\documentclass[superscriptaddress,
twocolumn,showpacs,preprintnumbers,amsmath,amssymb]{revtex4}
\usepackage{graphicx}
\usepackage{dcolumn}
\usepackage{bm}
\usepackage{latexsym}
\begin{document}
\title{Fluctuations in protein synthesis from a single RNA template:\\ 
stochastic kinetics of ribosomes} 
\author{Ashok Garai{\footnote{Corresponding author(E-mail: garai@iitk.ac.in)}}  }
\affiliation{Department of Physics, Indian Institute of Technology,
Kanpur 208016, India.}
\author{Debashish Chowdhury}
\affiliation{Department of Physics, Indian Institute of Technology,
Kanpur 208016, India.}
\author{T. V. Ramakrishnan}
\affiliation{Department of Physics, Banaras Hindu University, Varanasi 221005, India.}
\affiliation{Department of Physics, Indian Institute of Science, Bangalore 560012, India.}
\affiliation{Department of Physics, Indian Institute of Technology,
Kanpur 208016, India.}
\pacs{87.16.af; 87.16.dj}
\date{\today}%
\begin{abstract}
Proteins are polymerized by cyclic machines called ribosome which use 
their messenger RNA (mRNA) track also as the corresponding template 
and the process is called translation. We explore, in depth and detail, 
the stochastic nature of the translation. We compute various distributions 
associated with the translation process; one of them, namely dwell time 
distribution, has been measured in recent single ribosome experiments 
(Wen et al.  Nature {\bf 452}, 598 (2008)). The form of this distribution 
predicted by our theory is consistent with that extracted from the 
experimental data. For our quantitative calculations, we use a model 
that captures both the mechano-chemistry of each individual ribosome 
as well as their steric interactions. We also demonstrate the effects 
of the sequence inhomogeneities of real genes on the fluctuations and 
noise in translation. In principle, our new predictions can be tested 
by carrying out {\it in-vitro} experiments.
\end{abstract}
\maketitle

A genetic message, chemically encoded in the DNA, is first {\it 
transcribed} into a messenger RNA (mRNA) from which it is then 
{\it translated} into proteins \cite{alberts}. Both mRNA and 
proteins are linear polymers of monomeric subunits called nucleotide 
and amino acid, respectively. The genetic code contained in the 
sequence of codons (triplets of nucleotides) on an mRNA is translated 
into the corresponding  sequence of amino acids by a macromolecular 
machine, called ribosome \cite{spirin}. A ribosome is a cyclic machine. 
Each mechano-chemical cycle of this machine consists of several steps 
which result in the translocation of the ribosome by one codon on the 
mRNA template and the elongation of the protein by one amino acid. 
Thus, the mRNA template also serves as the track for motor-like movement 
of the ribosome during translation \cite{hill69,cross97}. In fact, 
a ribosome is like a mobile ``workshop'' which moves on an mRNA track 
and provides a platform where a coordinated action of many devices 
take place for the synthesis of each of the proteins.

Only a few papers over the last few years have reported results of 
single-ribosome imaging and manipulation 
\cite{blanchard04,uemura07,munro08,vanzi03,wang07,wen08}. 
These experiments have established that in each mechano-chemical 
cycle, the dwell time of a ribosome at any codon is random. Moreover, 
this dwell time is a sum of two time intervals, namely, (i) the 
duration for which it makes a mechanical pause and (ii) the time it 
takes to translocate to the next codon.

In this letter we report our theoretical results on the dwell time 
distribution \cite{wen08}, which characterizes the stochastic 
translocation-and-pause dynamics of the ribosomes. We also introduce 
a few new statistical distributions which characterize some other 
aspects of the stochastic nature of translation. We compute all 
these statistical distributions by carrying out computer simulations 
of a model of protein synthesis that captures both the 
mechano-chemistry of each individual ribosome, as it moves on 
the mRNA template, as well as their in-situ steric interactions 
\cite{basuchow}. To our knowledge, none of the earlier models of 
``ribosome-traffic''  
\cite{macdonald68,macdonald69,lakatos03,shaw03,shaw04a,shaw04b,chou03,chou04,schonherr1,schonherr2,dong}
have been used so far to investigate any of these quantitative measures 
of the stochastic kinetics of ribosomes. We treat the widths of these 
distributions as quantitative measures of ``translational noise'' 
arising from a single mRNA template. Moreover, in this letter, we 
demonstrate the effects of the heterogeneity of the codon sequence of 
real genes on this ``translational noise''.

We represent the single-stranded mRNA chain by a one-dimensional lattice 
where each site corresponds a single codon (triplet of nucleotides). 
The sites $i = 1$ and $i=L$ 
represent the start codon and stop codon, respectively. Each ribosome 
covers ${\ell}$ sites (i.e., ${\ell}$ codons) at a time; no lattice 
site is allowed to be covered simultaneously by more than one overlapping 
ribosome because of their steric exclusion. Irrespective of the length 
$\ell$, each ribosome moves forward by only one site in each step as 
it must translate successive codons one by one. We denote the position 
of a ribosome by the integer index of the leftmost lattice site it covers. 

\begin{figure}
\includegraphics[angle=-90,width=0.475\textwidth]{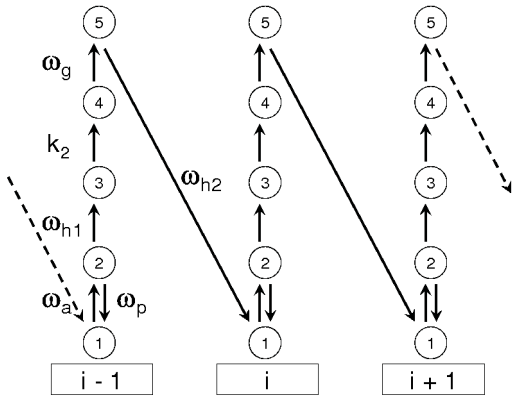}
\caption{A schematic representation of the biochemical cycle of a
single ribosome during the elongation stage of translation in our
model \cite{basuchow}. Each circle labelled by an integer index 
represents a distinct state in the mechano-chemical state of a 
ribosome. The index below the box labels the codon on the mRNA 
with which the ribosome binds. The symbols accompanied by 
the arrows define the rate constants for the corresponding 
transitions from one state to another. 
}
\label{states}
\end{figure}

\begin{figure} 
\begin{center}
\includegraphics[width=0.475\textwidth]{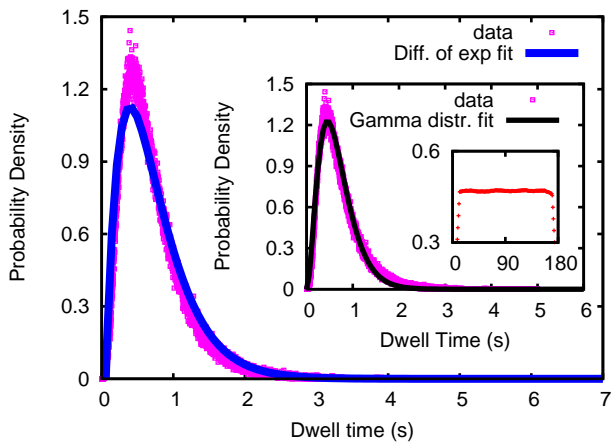}
\end{center}
\caption{Probability distribution of the dwell times of the ribosomes  
for a hypothetical homogeneous mRNA template for ${\ell} = 12$. The 
same set of data fits equally well with difference of two exponentials 
and with a gamma distribution (shown in the inset) Inset of the inset 
shows the {\it coverage density} profile of the ribosomes on the mRNA. 
The parameters are $\omega_{a} = \omega_{g} = 25$ s$^{-1}$, 
$\alpha = 0.0001$ s$^{-1}$. 
}
\label{fig-dwell}
\end{figure}

\begin{figure} 
\begin{center}
\includegraphics[width=0.475\textwidth]{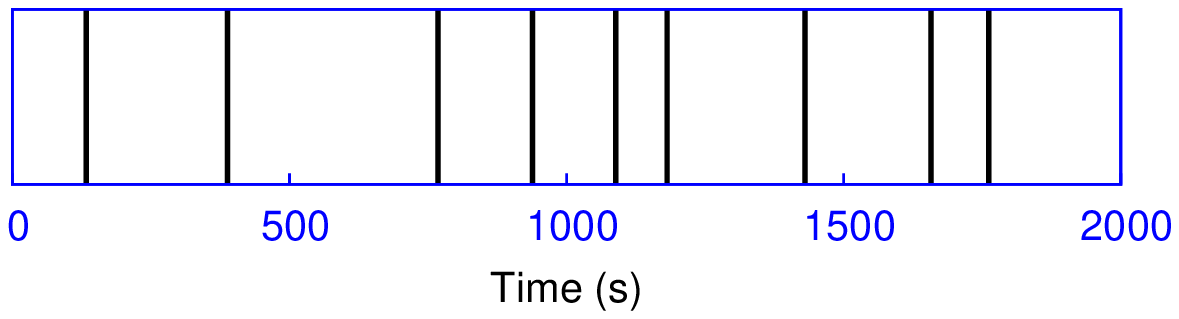}
\includegraphics[width=0.475\textwidth]{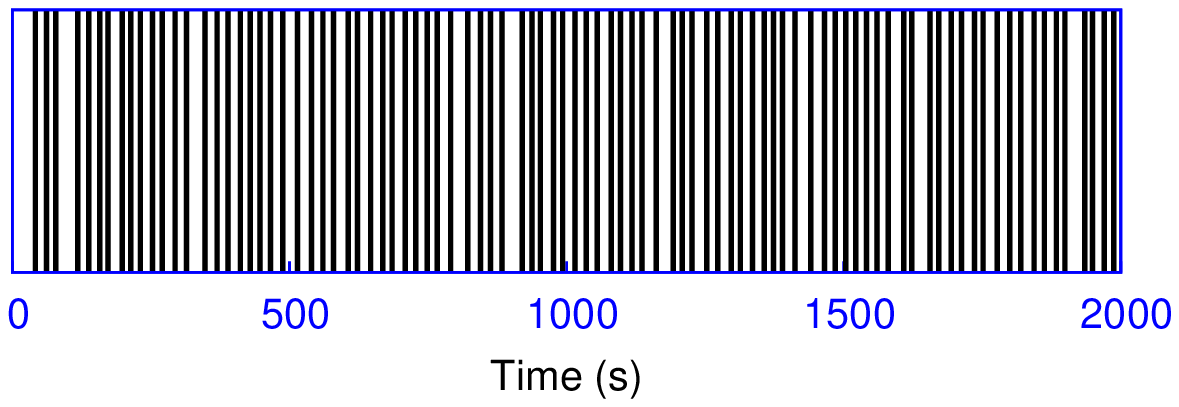}
\end{center}
\caption{Typical time series of the translation events for 
(a) crr gene of {\it Escherichia coli} K-12 strain MG1655 and 
(b) the corresponding hypothetical homogeneous mRNA template, 
both corresponding to $\omega_{a} = 2.5$s$^{-1}$, 
$\omega_{g} = 2.5$s$^{-1}$, $\omega_{h} = 10$s$^{-1}$ and 
$\alpha = 0.1$ s$^{-1}$.  
}
\label{fig-tseries}
\end{figure}

\begin{figure} 
\begin{center}
\includegraphics[width=0.475\textwidth]{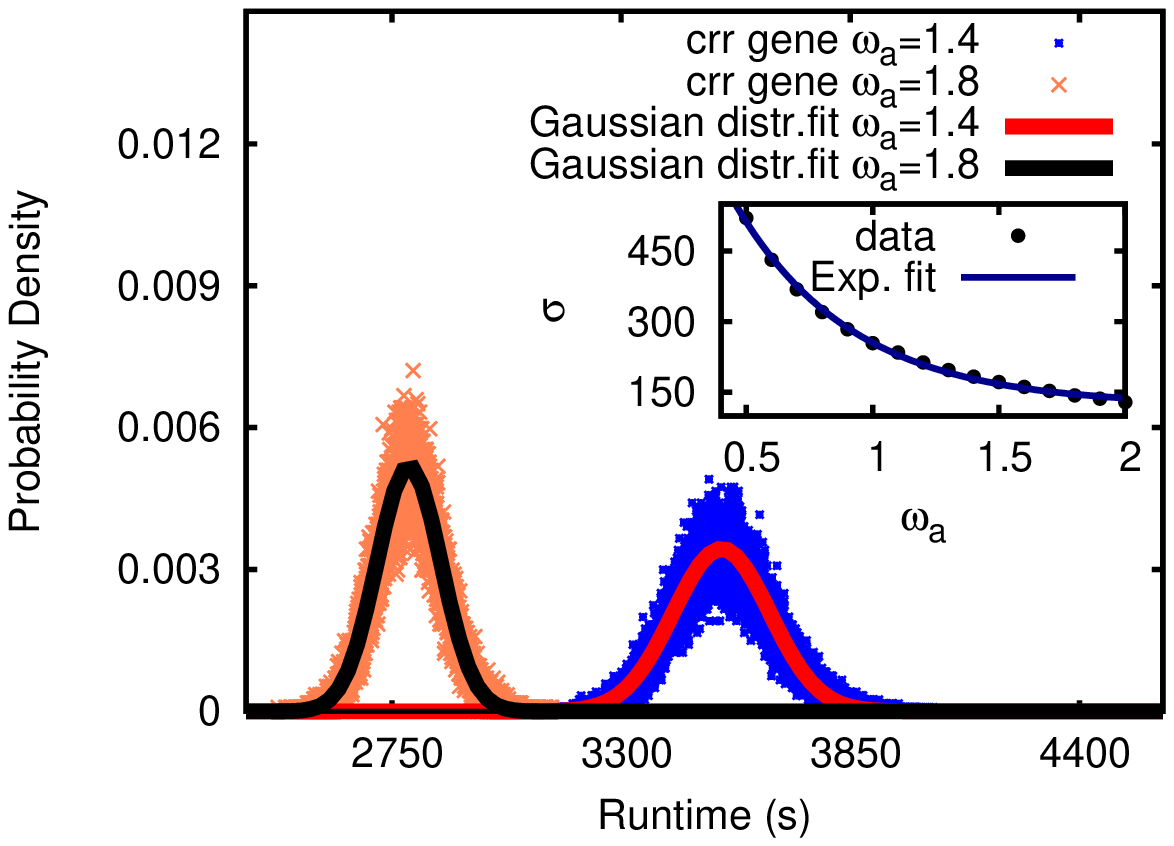}
\includegraphics[angle=-90,width=0.475\textwidth]{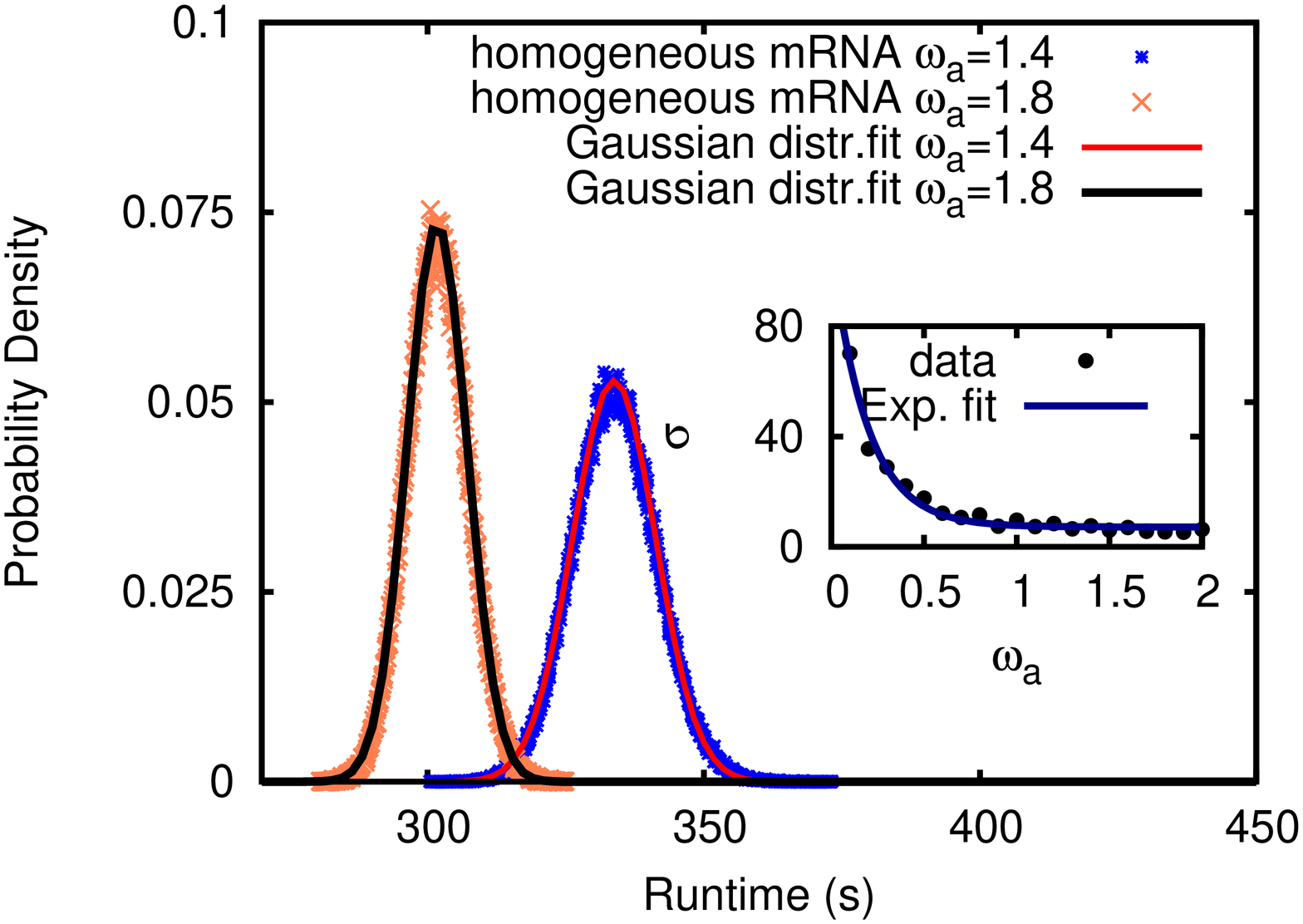}
\end{center}
\caption{Probability distribution of the times taken to complete 
the synthesis of a single polypeptide (which is identical to the 
probability distribution of the run times of ribosomes) for 
(a) crr gene of {\it Escherichia~coli} K-12 strain MG1655, 
and (b) the corresponding hypothetical homogeneous mRNA template. 
Both in (a) and (b), different curves correspond to different 
values of $\omega_{a}$, all for ${\ell} = 12$. The discrete data 
points have been obtained from our computer simulations of the 
model whereas the lines denote the {\it gaussian} best fits to 
these data. The insets show the exponential decrease of the 
corresponding noise strengths with $\omega_{a}$. In both (a) and (b), 
$\omega_{g} = 2.5$ s$^{-1}$ and $\alpha = 0.1$ s$^{-1}$.
}
\label{fig-rntmwa}
\end{figure}

\begin{figure} 
\begin{center}
\includegraphics[width=0.475\textwidth]{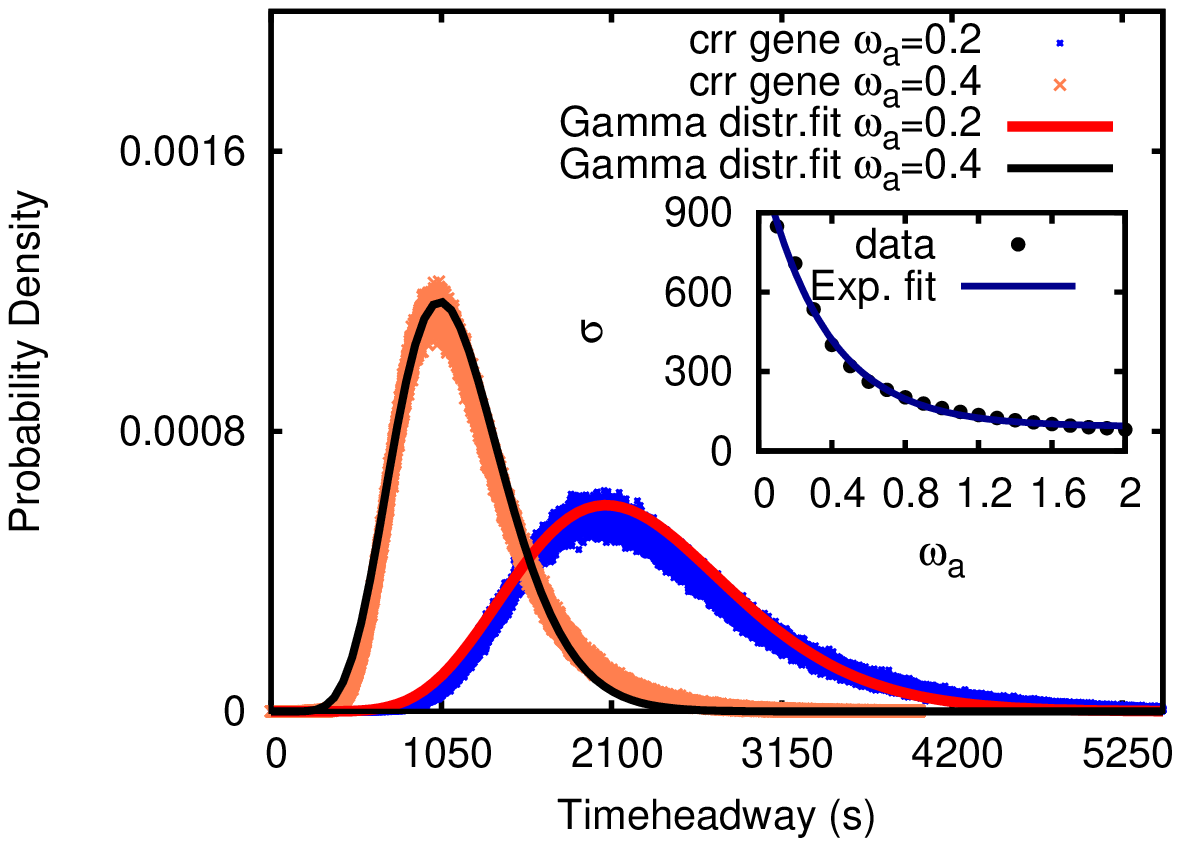}
\includegraphics[width=0.475\textwidth]{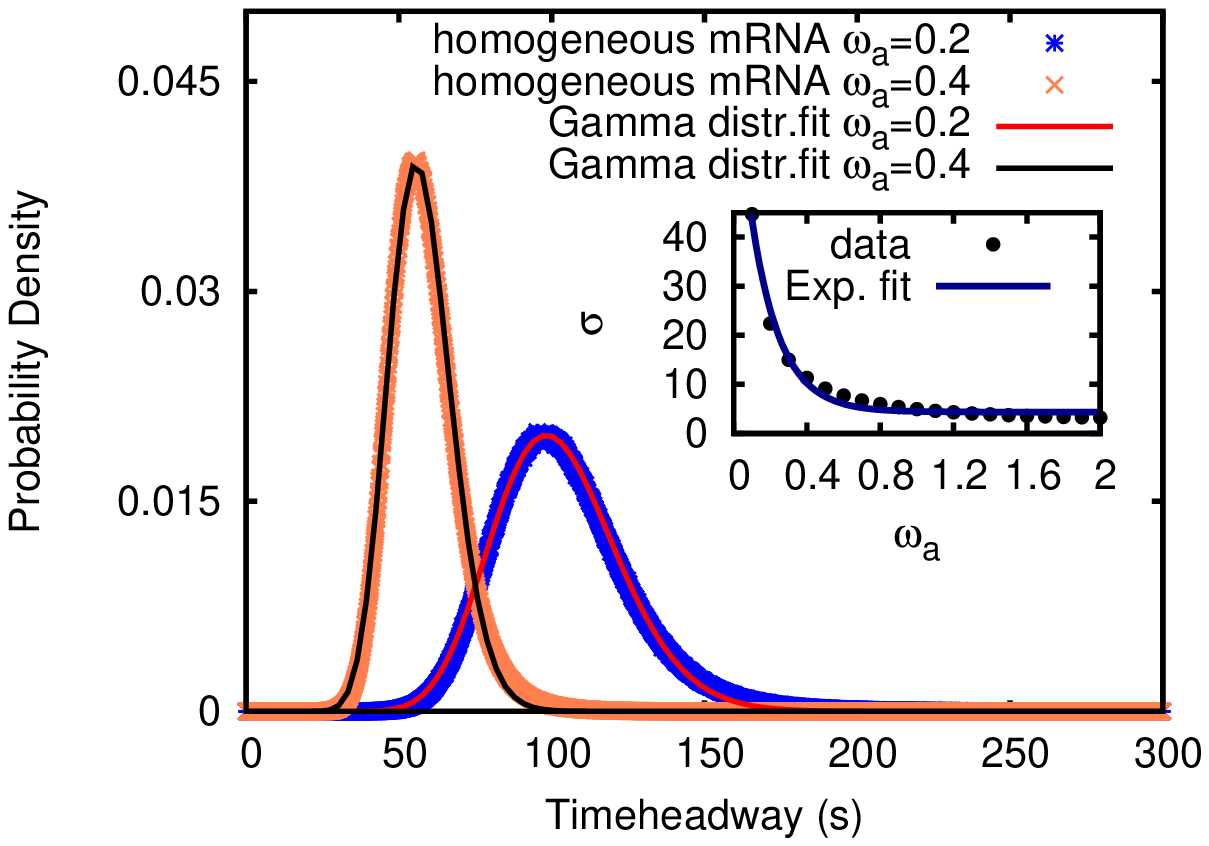}
\end{center}
\caption{Probability distribution of the time gaps between the 
completions of the synthesis of a successive polypeptides (which 
is identical to the probability distribution of the time-headways 
in the ribosome traffic) for 
(a) crr gene of {\it Escherichia~coli} K-12 strain MG1655, and 
(b) the corresponding hypothetical homogeneous mRNA template. 
Both in (a) and (b) different curves correspond to different 
values of   $\omega_{a}$, all for ${\ell} = 12$. The discrete 
data points have been obtained from our computer simulations of 
the model whereas the lines denote the {\it gamma} distributions 
fitted to these data. The insets show the exponential decrease of 
the corresponding noise strengths with $\omega_{a}$. In both (a) 
and (b), $\omega_{g} = 2.5$ s$^{-1}$ and $\alpha = 0.1$ s$^{-1}$.
}
\label{fig-tmgpwa}
\end{figure}

The fig.\ref{states} captures the mechano-chemical cycle of each ribosome 
in the stage of elongation of the protein. The arrival of the correct 
amino-acid (bound to an adapter molecule called tRNA) and its 
recognition by the ribosome located at the site $i$ triggers transition 
from the chemical state 1 to 2 in the same location. The transition 
from state 2 to state 3 is driven by hydrolysis of GTP. Departure of 
the phosphate group, which is one of the products of GTP hydrolysis, 
results in the intermediate state 4. The peptide bond formation between 
the growing protein and the incoming amino acid monomer (and some 
associated biochemical processes), which leads to the elongation of 
the protein by one amino acid monomer, is captured by the next 
transition to the state 5. All the subsequent processes, including 
hydrolysis of another GTP molecule, the forward translocation of 
the ribosome by one codon and the departure of a naked tRNA from the 
ribosome complex are captured by a single effective transition 
from state 5 at site $i$ to the state $1$ at the site $i+1$. More 
detailed explanations of the states and the transitions are given in 
ref.\cite{basuchow}.

The average number of ribosomes crossing the stop codon, per unit time, 
on the template mRNA is called the {\it flux} of ribosomes. The average 
rate of elongation of a protein is proportional to the average velocity 
of a ribosome and, therefore, the flux is a measure of the total rate of 
synthesis of the protein encoded by the mRNA on which the ribosomes 
move. The flux and the average density profiles of the ribosomes on 
the mRNA track in our model have been reported in ref.\cite{basuchow}.


The time interval $t_{d}$ between the arrival of a ribosome at a 
specific codon and its subsequent departure from there is defined as 
the dwell time at that codon. The run time $T$ of a ribosome is the 
time is takes to run from the start codon to the stop codon on the 
mRNA. In other words, $T$ is the time taken by a ribosome to 
synthesize a single protein. Similarly, following the terminology of 
traffic science \cite{css}, we identify the time interval between 
the departure of the successive ribosomes from the stop codon as the 
time-headway $\tau$. Equivalently, $\tau$ is the time interval in 
between the completion of the synthesis of successive proteins from 
the same mRNA template. 

In this letter we compute the distributions $P(t_{d})$, ${\tilde{P}}(T)$,  
and ${\cal P}(\tau)$ of the probabilities of $t_{d}$, $T$ and $\tau$. 
We treat the fluctuations, i.e., root-mean-square (rms) deviations, of 
$t_{d}, T$ and $\tau$ as quantitative measures of noise in the 
translation of a single mRNA. 
Analogous measures of transcriptional noise have been introduced 
recently to characterize the stochasticity of polymerization of RNA 
molecules from a DNA template \cite{tripathi}.

All the calculations reported in this paper have been obtained by 
imposing {\it open} boundary conditions which mimics protein synthesis 
more realistically. The symbols $\alpha$ and $\beta$ denote the 
probabilities of attachment and detachment, respectively, in time 
$\Delta t$. So, the probability of attachment per unit time (which 
we call $\omega_{\alpha}$) is the solution of the equation 
$\alpha{}=1-e^{-\omega_{\alpha}\times{}\Delta t}$ (in all our 
numerical calculations we take $\Delta t = 0.001$ s). Similarly, we 
define the corresponding parameter $\omega_{\beta}$ for termination.  
For the same reasons as elaborated in ref.\cite{basuchow}, we assume 
that $\omega_{h1} \simeq \omega_{h2} = \omega_{h}$. Moreover, 
throughout this letter we use $\omega_{h} = 10$ s$^{-1}$, 
$\omega_p = 0.0028$ s$^{-1}$, $k_{2} = 2.4$ s$^{-1}$ and 
$\beta = 1$ s$^{-1}$ which were used in ref.\cite{basuchow} for the 
bacteria {\it E-coli}; the values of the other parameters will be 
given in the appropriate figure captions. We incorporate the effects 
of the inhomogeneity of the sequence of codons in the crr gene of 
{\it Escherichia~coli} K-12 strain MG1655 \cite{databank} in our 
model exactly the same way as it was done in ref.\cite{basuchow}.

Because of the intrinsic stochasticity of the steps of the 
mechano-chemical cycle of the ribosome, the dwell time fluctuates 
even if all the codons on the mRNA track are identical. A typical 
distribution of the dwell times of the ribosomes during the 
translation of a hypothetical homogeneous mRNA is shown in 
fig.\ref{fig-dwell}. The numerical data obtained from computer 
simulations of our model can be fitted to the difference of two 
exponentials; this is consistent with the corresponding recent 
experimental observation \cite{wen08}. However, the same data fits 
with a gamma distribution almost equally well (see the inset of 
fig.\ref{fig-dwell}).

Typical time series of the translation events is shown in 
fig.\ref{fig-tseries} for the crr gene of {\it Escherichia~coli}
K-12 strain MG1655 together with a time-series for the 
corresponding homogeneous mRNA template where all the rate 
constants other than $\omega_{a}$ are same. The longer gaps 
between the events for the real gene arises from the fact that 
a ribosome has to wait for long periods at the ``hungry codons'' 
\cite{basuchow}.

We have plotted the distribution $\tilde{P}(T)$ for the crr gene of 
the {\it Escherichia~coli} K-12 strain MG1655, for different values 
of the model parameters $\omega_{a}$ in fig.\ref{fig-rntmwa}; 
the data for the corresponding hypothetical homogeneous mRNA template 
are plotted in fig.\ref{fig-rntmwa}(b).  
In fig.\ref{fig-tmgpwa}  we have plotted the corresponding data for 
${\cal P}_{\tau}$. The variation of the strength of the noise with 
the model parameters are shown in the insets of the respective figures. 
Both the measures of translational noise fall exponentially with the 
increase of $\omega_{a}$. In other words, increase in the availability 
of the monomeric subunits (which is indicated by $\omega_{a}$) reduce 
the noise level. Similar trend of variation of noise with $\omega_{h}$ 
(i.e., the rate of ``fuel'' consumption) has been observed (but not 
shown graphically). 

Comparing the data in fig.\ref{fig-rntmwa}(a) and fig.\ref{fig-rntmwa}(b)  
we conclude that the sequence inhomogeneity of real genes not only 
slows down the polymerization of the proteins, but also makes the 
process more noisy as compared to the translation of the hypothetical 
homogeneous gene. Similarly, comparing the data in fig.\ref{fig-tmgpwa}(a) 
with those in fig.\ref{fig-tmgpwa}(b) we establish that sequence 
inhomogeneity of real genes leads to longer mean, as well as stronger 
fluctuations, in $\tau$ than for the hypothetical homogeneous template.

The data for $\tilde{P}(T)$, obtained from computer simulations, fit 
well with a {\it gaussian} distribution 
In contrast, the best fit to those for ${\cal P}_{\tau}$ is a {\it gamma} 
disribution 
Such long-tail distributions are quite common in gene expression 
and describe the characteristic features of various statistical 
properties of gene expression \cite{xie06c,morelli07,krishna05}.

In this letter we have developed a new conceptual framework for analyzing 
the intrinsic stochasticity in the process of polymerization of proteins 
by ribosome machines from a single mRNA template. The widths of the 
statistical distributions, which characterize different aspects of this 
stochasticity, serve as quantitative measures of noise in the translation 
of a single mRNA. By comparing our results for a specific gene of the 
bacteria {\it Escherichia~coli} with those for the corresponding artificial 
homogeneous mRNA template, we have demonstrated the effects of the 
sequence inhomogeneities of real genes on the translational noise. 
The nature of the dwell time distributions predicted by our theory is 
consistent with the corresponding observations \cite{wen08} in recent 
single-ribosome experiments. We hope our other predictions will 
stimulate new experiments on translational noise.

We thank Aakash Basu for his help in our numerical calculations during 
the initial stages of this work. This work has been supported (through 
DC) by Council of Scientific and Industrial Research (CSIR), government 
of India.



\begin{thebibliography}{99}
                                                                                
\bibitem{alberts} B. Alberts et al. 
{\sl Molecular Biology of the Cell}, (4th edition)
(Garland Publishing, 2002).

\bibitem{spirin} A. S. Spirin, {\it Ribosomes}, (Springer, 2000); 
FEBS Lett. {\bf 514}, 2 (2002). 
\bibitem{hill69} T.L. Hill. Proc. Natl. Acad. Sci. {\bf 64}, 267 (1969).
\bibitem{cross97} R. A. Cross, Nature {\bf 385}, 18 (1997).

\bibitem{macdonald68} C. MacDonald, J. Gibbs and A. Pipkin, Biopolymers,
{\bf 6}, 1 (1968).
\bibitem{macdonald69} C. MacDonald and J. Gibbs, Biopolymers, {\bf 7},
707 (1969).
\bibitem{lakatos03} G. Lakatos and T. Chou, J. Phys. A {\bf 36}, 2027
(2003).
\bibitem{shaw03} L.B. Shaw, R.K.P. Zia and K.H. Lee, Phys. Rev. E
{\bf 68}, 021910 (2003).
\bibitem{shaw04a} L.B. Shaw, J.P. Sethna and K.H. Lee, Phys. Rev. E
{\bf 70}, 021901 (2004).
\bibitem{shaw04b} L.B. Shaw, A.B. Kolomeisky and K.H. Lee, J. Phys. A
{\bf 37}, 2105 (2004).
\bibitem{chou03} T. Chou, Biophys. J., {\bf 85}, 755 (2003).
\bibitem{chou04} T. Chou and G. Lakatos, Phys. Rev. Lett. {\bf 93},
198101 (2004).
\bibitem{schonherr1} G. Sch\"onherr and G. M. Sch\"utz, J. Phys. A {\bf 37}, 8215 (2004).
\bibitem{schonherr2} G. Sch\"onherr, Phys. Rev. E {\bf 71}, 026122 (2005).
\bibitem{dong} J.J. Dong, B. Schmittmann and R.K.P. Zia, J. Stat. Phys.
{\bf 128}, 21 (2007).


\bibitem{basuchow} A. Basu and D.Chowdhury, Phys. Rev. E {\bf 75}, 021902 (2007).

\bibitem{blanchard04} S. Blanchard et al., 
Nat. Str. \& Mol. Biol. {\bf 11}, 1008 (2004). 
\bibitem{uemura07} S. Uemura et al., 
Nature {\bf 446}, 454 (2007). 
\bibitem{munro08} J.B. Munro et al., 
Biopolymers {\bf 89}, 565-577 (2008).
\bibitem{vanzi03} F. Vanzi et al., 
RNA, {\bf 9}, 1174-1179 (2003).
\bibitem{wang07} Y. Wang et al., 
Biochemistry {\bf 46}, 10767-10775 (2007).
\bibitem{wen08} J.D. Wen et al., 
Nature {\bf 452}, 598 (2008).

\bibitem{css} D. Chowdhury, L. Santen and A. Schadschneider, Phys. Rep. 
{\bf 329}, 199 (2000).
\bibitem{derrida} B. Derrida and M.R. Evans, in: {\it Non-equilibrium
Statistical Mechanics in One Dimension}, ed. V. Privman (Cambridge
University Press, 1997); B. Derrida, Phys. Rep. {\bf 301}, 65 (1998)
\bibitem{schuetz} G. M. Sch\"utz, {Phase Transitions and Critical
Phenomena}, vol. 19 (Acad. Press, 2001).
\bibitem{tripathi} T. Tripathi and D. Chowdhury, Phys. Rev. E {\bf 77}, 
011921 (2008). 
\bibitem{databank} http://www.genome.wisc.edu/sequencing/k12.htm
\bibitem{xie06c} N. Friedman, L. Cai and X. S. Xie, Phys. Rev. Lett. 
{\bf 97}, 168302 (2006).
\bibitem{morelli07} L.G. Morelli and F. J\"ulicher, Phys. Rev. Lett. 
{\bf 98}, 228101 (2007). 
\bibitem{krishna05} S. Krishna et al., 
Proc. Natl. Acad. Sci. USA {\bf 102}, 4771 (2005).

\end{thebibliography}
\end{document}